\documentclass[aps,preprint, floatfix]{revtex4}%
\usepackage{amsfonts}
\usepackage{amsmath}
\usepackage{amssymb}
\usepackage{graphicx}%
\begin{document}
\title[Microtubules Interacting with a Boundary]{Microtubules Interacting with a Boundary: Mean Length and Mean First-Passage Times}
\author{Bela M. Mulder}
\affiliation{FOM Institute AMOLF, Science Park 104, 1098XG Amsterdam, the Netherlands}
\keywords{microtubules; boundary collisions; mean first-passage time}

\begin{abstract}
We formulate a dynamical model for microtubules interacting with a
catastrophe-inducing boundary. In this model microtubules are either waiting
to be nucleated, actively growing or shrinking, or stalled at the boundary. We
first determine the steady-state occupation of these various states and the
resultant length distribution. Next, we formulate the problem of the Mean
First-Passage Time to reach the boundary in terms of an appropriate set of
splitting probabilities and conditional Mean First-Passage Times, and solve
explicitly for these quantities using a differential equation approach. As an
application, we revisit a recently proposed search-and-capture model for the
interaction between microtubules and target chromosomes [Gopalakrishnan \&
Govindan, Bull. Math. Biol. 73:2483--506 (2011)]. We show how our approach
leads to a direct and compact solution of this problem.

\end{abstract}
\date{January 25, 2012}
\revised[Revised: ]{\today}

\maketitle

\section{Introduction}

Microtubules (hereafter abbreviated as MTs) are filamentous macro-polymers
built from tubulin dimers. They are one of the components of the cytoskeleton
of all eukaryotic cells. They play a number of roles ranging from providing
mechanical stability to the cell, serving as transport pathways enabling
linear transport of versicular cargo by motor proteins, and providing forces
for the positioning of organelles and other intracellular components (for a
general overview see \cite{Alberts2002}). Perhaps their most striking function
appears during mitosis, where they form the mitotic spindle, the machinery for
positioning and separating the duplicated chromosomes prior to cell division.
They owe their functional plasticity to an intrinsic so-called `dynamical
instability' mechanism \cite{Mitchison1984} that causes individual MTs to
stochastically alternate between growing and shrinking states. By controlling
this dynamical process, through MT associated proteins (MAPS) that nucleate
new MTs or selectively stabilize or destabilize them by locally or globally
changing the rates with which they switch between dynamical states, cells are
able to reconfigure MT assemblies on timescales as fast as a few minutes. The
canonical model to describe MT dynamics was developed in the early 1990's by
Dogterom and Leibler \cite{Dogterom1993}. This model showed that isolated MTs,
depending on their dynamical parameters, can either be in a regime of bounded
growth leading to an exponential length distribution in a steady state, or in
a regime of unbounded growth in which the MT length on average increases
linearly in time. Of course, MTs `live' within the confines of a finite size
cell, whose dimensions ($\sim10\mu m$) are comparable to the observed lengths
of MTs. Interactions between MTs and boundaries, be it the cell cortex or the
surface of other intracellular compartments, are therefore important. Indeed,
a number of MT functions depends critically on these interactions: examples
are nuclear positioning in yeast \cite{Tran2001}, spindle positioning in
\emph{C. elegans }\cite{Grill2003}, and the orientation of the cortical MT
array in plant cells \cite{Ambrose2011}.

In spite of this clear relevance, it appears that a systematic approach to the
theory of MTs interacting with boundaries is lacking from the literature. The
one problem of this type which did receive substantial attention, is the
search-and-capture mechanism by which MTs are thought to find the condensed
chromosomes prior to mitosis \cite{Holy1994}\cite{Holy1997}\cite{Wollman2005}%
\cite{Gopalakrishnan2011}, which involves estimating the mean first-passage
time of a MT to hit a limited size target at a distance from its nucleation
point. However, although these works in fact do contain some of the basic
features of the MT-boundary problem, it is mostly hidden (literally in the
case of Ref. \cite{Holy1997}, actually an unpublished thesis) under the
specifics of the intended application. Moreover, these works also rely heavily
on `forward' techniques involving the time evolution of the full probability
density for reaching a given state from specified initial conditions. Although
this approach obviously yields a full solution of the problem, for
passage-time problems, which effectively require integrating over final
states, the full probability density is in a sense a form of \textquotedblleft
overkill\textquotedblright. The treatment of this type of problems can in fact
be simplified considerably by using `backward' techniques, as is
e.g.\ elegantly illustrated for diffusion problems in Redner's monograph
\cite{Redner2001}. In the present work we show how this approach can be used
from the ground up to solve the problem of a MT interacting with a boundary.

The outline of the paper is as follows: In Section \ref{sec:meanl} we
introduce the dynamical model of a MT interacting with a boundary, solve for
its steady state properties, such as the average length, and choose an
appropriate set of non-dimensionalized parameters and variables. In Section
\ref{sec:MFPT} we turn to the analysis of the mean first-passage time to the
boundary, formally solving this in terms of a small set of splitting
probabilities and conditioned mean first-passages times, which are
subsequently determined explicitly through the solution of appropriate linear
boundary value problems. We then use some biological data on MT dynamics to
estimate order of magnitudes for the quantities involved. Finally, in Section
\ref{sec:GG} we revisit the recent search-and-capture model discussed by
Gopalakrishnan and Govindan \cite{Gopalakrishnan2011}, and show how it is
compactly solved using the techniques introduced. We then finish with a number
of concluding remarks and two technical appendices.

\section{Mean length\label{sec:meanl}}

\subsection{Dynamical model}

\begin{figure}[ptb]
\includegraphics[width=\textwidth]{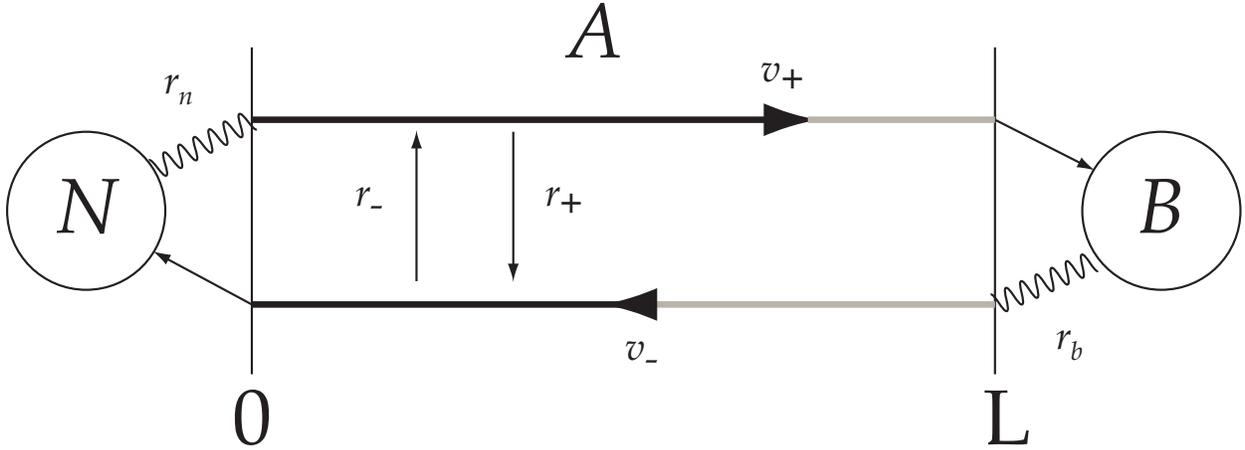}\caption{Schematic of our model
showing the three components of the state space: The nucleation state $N$ from
which MTs are nucleated at rate $r_{n}$. The active state $A$, in which MTs
either grow or shrink, with velocities $v_{+}$ and $v_{-}$, and switch between
these two states with rates $r_{+}$ and $r_{-}$, and a barrier state $B$ from
which MTs exit at a rate $r_{b}$.}%
\label{fig:barrier}%
\end{figure}

The standard two-state dynamical instability model describes MTs with length
$l$ that are either growing with velocity $v_{+}$ or shrinking with velocity
$v_{-}$ and can switch between the growing and the shrinking state (a
\emph{catastrophe}) with a constant rate $r_{+}$ and between the shrinking and
the growing state (a \emph{rescue}) with a constant rate $r_{-}$. Collectively
we call these two states the \emph{active} states, and denote the
corresponding state space by $A$. We extend this model by two more states: a
\emph{nucleation} state $N$, in which a MT enters upon shrinking back to zero
length, and from which it can be (re)nucleated into a zero-length growing
state at a constant rate $r_{n}$, and a boundary state $B$, in which a MT
enters upon hitting a boundary at a distance $L$ from the nucleation point,
and which it leaves in a shrinking state at a rate $r_{b}$. Formally the state
space of this extended model is therefore given by $\Omega=N\cup A\cup B$. We
illustrate the model and its state space in Figure~\ref{fig:barrier}.
It should be noted that, from a biophysical point of view, the
model is of course an idealization. In reality a growing MT impinging on a
boundary will generate forces. These forces will affect the growth velocity
and the propensity to switch to the shrinking state, so that the latter is no
longer a simple Poisson process \cite{Janson2003}. Moreover, these forces may
also deform the boundary, lead to buckling of the MT, or cause it to slide
along the boundary (for a review see \cite{Dogterom2005}). These additional
complexities, however, will have limited impact on the results to be presented
here, as long as the residence time at the boundary is small compared to the
time to traverse the distance between nucleation point and boundary, which is
certainly the case for effectively reflective boundaries for which $r_{b}\gg
r_{+}$, and we choose to ignore them here.

The dynamics of the model described above defines a time-homogeneous Markov
process on the full state space. It is, however, technically convenient to
split the dynamics on the `active' part of the state space $A$ from that on
the two `waiting' states $N$ and $B$, and deal with the communication between
these different states through boundary conditions. We therefore first define
the probability densities (per unit length) $m_{s}\left(  l,t|\omega_{0}%
,t_{0}\right)  $ for an active MT to have length $l$ and be in state $s=+,-$
at time $t$, given that it was in some state $\omega_{0}$ at time $t_{0}<t$.
These densities satisfy the evolution equations%
\begin{align}
\frac{\partial}{\partial t}m_{+}\left(  l,t|\omega_{0},t_{0}\right)   &
=-v_{+}\frac{\partial}{\partial l}m_{+}\left(  l,t|\omega_{0},t_{0}\right)
-r_{+}m_{+}\left(  l,t|\omega_{0},t_{0}\right)  +r_{-}m_{-}\left(
l,t|\omega_{0},t_{0}\right)  ,\label{eq:actp}\\
\frac{\partial}{\partial t}m_{-}\left(  l,t|\omega_{0},t_{0}\right)   &
=v_{-}\frac{\partial}{\partial l}m_{-}\left(  l,t|\omega_{0},t_{0}\right)
-r_{-}m_{-}\left(  l,t|\omega_{0},t_{0}\right)  +r_{+}m_{+}\left(
l,t|\omega_{0},t_{0}\right)  . \label{eq:actm}%
\end{align}
Likewise we define the probability $M_{n}\left(  t|\omega_{0},t_{0}\right)  $
for the MT to be in the nucleation state at time $t$, given that it was in
some state $\omega_{0}\in\Omega$ at time $t_{0}<t$. This probability satisfies
the evolution equation%
\begin{equation}
\frac{\partial}{\partial t}M_{n}\left(  t|\omega_{0},t_{0}\right)  =v_{-}%
m_{-}\left(  0,t|\omega_{0},t_{0}\right)  -r_{n}M_{n}\left(  t|\omega
_{0},t_{0}\right)  . \label{eq:0}%
\end{equation}
The probability $M_{b}\left(  t|\omega_{0},t_{0}\right)  $ for the MT to be in
the boundary state at time $t$ in turn satisfies%
\begin{equation}
\frac{\partial}{\partial t}M_{b}\left(  t|\omega_{0},t_{0}\right)  =v_{+}%
m_{+}\left(  L,t|\omega_{0},t_{0}\right)  -r_{b}M_{b}\left(  t|\omega
_{0},t_{0}\right)
\end{equation}
This system of equations is closed by the boundary conditions%
\begin{align}
v_{+}m_{+}\left(  0,t|\omega_{0},t_{0}\right)   &  =r_{n}M_{n}\left(
t|\omega_{0},t_{0}\right)  ,\\
v_{-}m_{-}\left(  L,t|\omega_{0},t_{0}\right)   &  =r_{b}M_{b}\left(
t|\omega_{0},t_{0}\right)  .
\end{align}
By construction the dynamics on the full state space conserves probability,
and indeed if we define the total probability%
\begin{align}
M\left(  t|\omega_{0},t_{0}\right)   &  =M_{n}\left(  t|\omega_{0}%
,t_{0}\right)  +M_{a}\left(  t|\omega_{0},t_{0}\right)  +M_{b}\left(
t|\omega_{0},t_{0}\right) \nonumber\\
&  =M_{n}\left(  t|\omega_{0},t_{0}\right)  +\int_{0}^{\infty}dl\,\left\{
m_{+}\left(  l,t|\omega_{0},t_{0}\right)  +m_{-}\left(  l,t|\omega_{0}%
,t_{0}\right)  \right\}  +M_{b}\left(  t|\omega_{0},t_{0}\right)
\label{eq:norm}%
\end{align}
then%
\begin{equation}
\frac{\partial}{\partial t}M\left(  t|\omega_{0},t_{0}\right)  =0,
\end{equation}
allowing us to set $M\left(  t|\omega_{0},t_{0}\right)  =1$.

\subsection{Steady state behaviour}

\label{sec:SS} In steady state the probabilities do not depend on time nor on
initial conditions, allowing us to write the evolution equations as a set of
balance equations%
\begin{align}
v_{+}\frac{d}{dl}m_{+}\left(  l\right)   &  =-r_{+}m_{+}\left(  l\right)
+r_{-}m_{-}\left(  l\right)  ,\label{eq:SSp}\\
-v_{-}\frac{d}{dl}m_{-}\left(  l\right)   &  =-r_{-}m_{-}\left(  l\right)
+r_{+}m_{+}\left(  l\right)  ,\label{eq:SSm}\\
v_{-}m_{-}\left(  0\right)   &  =r_{n}M_{n},\label{eq:SSn}\\
v_{+}m_{+}\left(  L\right)   &  =r_{b}M_{b}, \label{eq:SSb}%
\end{align}
to be supplemented by the boundary conditions%
\begin{align}
v_{+}m_{+}\left(  0\right)   &  =r_{n}M_{n},\label{eq:SSBC0}\\
v_{-}m_{-}\left(  L\right)   &  =r_{b}M_{b}. \label{eq:SSBCL}%
\end{align}
Adding Eqs.~(\ref{eq:SSp}) and (\ref{eq:SSm}) yields%
\begin{equation}
\frac{d}{dl}\left\{  v_{+}m_{+}\left(  l\right)  -v_{-}m_{-}\left(  l\right)
\right\}  =0.
\end{equation}
Combining either Eqs.~(\ref{eq:SSn}) and (\ref{eq:SSBC0}), or
Eqs.~(\ref{eq:SSb}) and (\ref{eq:SSBCL}), shows that the constant of
integration vanishes, and hence%
\begin{equation}
v_{+}m_{+}\left(  l\right)  =v_{-}m_{-}\left(  l\right)  . \label{eq:SSbal}%
\end{equation}
This identity is now used to eliminate $m_{-}\left(  l\right)  $ from
(\ref{eq:SSp}) from which we then readily find that for $l\leq L$
\begin{align}
m_{+}\left(  l\right)   &  =\frac{r_{n}M_{n}}{v_{+}}e^{-l/\bar{l}},\\
m_{-}\left(  l\right)   &  =\frac{r_{n}M_{n}}{v_{-}}e^{-l/\bar{l}},
\end{align}
where the length%
\begin{equation}
\bar{l}=\left(  \frac{r_{+}}{v_{+}}-\frac{r_{-}}{v_{-}}\right)  ^{-1}%
\end{equation}
is of course only positive when $r_{+}v_{-}-r_{-}v_{+}>0$, the so called
bounded-growth regime, and represents the steady-state average length of an
active MT in the absence of the boundary. Although in the presence of
boundaries one can also consider the unbounded growth regime $r_{+}v_{-}%
-r_{-}v_{+}<0$, as was done e.g.\ in \cite{Govindan2004} for the case where
the boundary is fully reflecting, we will not do so here, and focus
exclusively on the bounded growth case. The dependence of the probability that
the MT is in the boundary state can, from (\ref{eq:SSb}), be shown to obey the
following relationship%
\begin{equation}
M_{b}=\frac{r_{n}}{r_{b}}e^{-\frac{L}{\bar{l}}}M_{n}%
\end{equation}
The normalization condition (\ref{eq:norm}) can then be used to determine the
probabilities for an MT to be in a nucleation state, an active state and a
boundary state respectively
\begin{align}
M_{n}  &  =\frac{1}{1+r_{n}\bar{l}\left(  \frac{1}{v_{+}}+\frac{1}{v_{-}%
}\right)  \left(  1-e^{-L/\bar{l}}\right)  +\frac{r_{n}}{r_{b}}e^{-L/\bar{l}}%
},\label{eq:MdaveB}\\
M_{a}  &  =\frac{r_{n}\bar{l}\left(  \frac{1}{v_{+}}+\frac{1}{v_{-}}\right)
\left(  1-e^{-L/\bar{l}}\right)  }{1+r_{n}\bar{l}\left(  \frac{1}{v_{+}}%
+\frac{1}{v_{-}}\right)  \left(  1-e^{-L/\bar{l}}\right)  +\frac{r_{n}}{r_{b}%
}e^{-L/\bar{l}}},\\
M_{b}  &  =\frac{\frac{r_{n}}{r_{b}}e^{-L/\bar{l}}}{1+r_{n}\bar{l}\left(
\frac{1}{v_{+}}+\frac{1}{v_{-}}\right)  \left(  1-e^{-L/\bar{l}}\right)
+\frac{r_{n}}{r_{b}}e^{-L/\bar{l}}}%
\end{align}
Taking the limit $L\rightarrow\infty$ we find $M_{a}\propto r_{n}\bar
{l}\left(  \frac{1}{v_{+}}+\frac{1}{v_{-}}\right)  $. As density $\propto$
nucleation rate $\times$ lifetime, this suggests that the time%
\begin{equation}
\bar{t}=\bar{l}\left(  \frac{1}{v_{+}}+\frac{1}{v_{-}}\right)  =\frac
{v_{+}+v_{-}}{r_{+}v_{-}-r_{-}v_{+}} \label{eq:tmean}%
\end{equation}
is the expected lifetime of a, otherwise unconstrained, length-zero newly
nucleated MT, a result indeed first derived by Rubin \cite{Rubin1988}.

We now define the mean length of the active MTs
\begin{equation}
\left\langle l\right\rangle _{a}=\frac{1}{M_{a}}\int_{0}^{L}dl\,l\left\{
m_{+}\left(  l\right)  +m_{-}\left(  l\right)  \right\}  =\bar{l}\frac{\left(
1-\left(  1+\frac{L}{\bar{l}}\right)  e^{-L/\bar{l}}\right)  }{\left(
1-e^{-L/\bar{l}}\right)  } \label{eq:laveB}%
\end{equation}
The time-averaged length of the MT over the full ensemble, is then simply%
\begin{equation}
\left\langle l\right\rangle =M_{a}\left\langle l\right\rangle _{a}+M_{b}L
\end{equation}
We can readily check the limits $\left\langle l\right\rangle _{a}%
\rightarrow\bar{l}$ when $L\rightarrow\infty$, and $\left\langle
l\right\rangle _{a}\simeq\frac{1}{2}L$ for $L\rightarrow0$. The latter limit
can be understood by considering that for very small $L$ the growing and
shrinking traversal times $L/v_{+}$ and $L/v_{-}$ become small with respect to
the mean time between between catastrophes, $1/r_{+}$, and rescues, $1/r_{-}$,
respectively, so that the MT is deterministically \textquotedblleft
bouncing\textquotedblright\ between the endpoints $l=0$ and $l=L$.

\subsection{Dimensional analysis}

\label{sec:dimen}We have deliberately deferred the dimensional analysis of the
system up to this point to allow the results of the steady state solution to
guide us to a natural choice of the units of length and time. In view of
(\ref{eq:MdaveB}) and (\ref{eq:laveB}), we choose $\bar{l}$ (\ref{eq:tmean})
as the unit length, and $\bar{t}$ as the unit of time. For completeness sake,
we can also introduce the unit of speed
\begin{equation}
\frac{1}{\bar{v}}\equiv\frac{\bar{t}}{\bar{l}}=\frac{1}{v_{+}}+\frac{1}{v_{-}}%
\end{equation}
By convention we will adopt the Greek alphabet do denote dimensionless
quantities. We can now introduce the dimensionless \emph{parameters}%
\begin{table}[ptb]%
\begin{tabular}
[c]{|l|l|l|}\hline
Symbol & Definition & Meaning\\\hline\hline
$\nu_{+}$ & $\frac{v_{+}}{\bar{v}}$ & growth speed\\\hline
$\nu_{-}$ & $\frac{v_{-}}{\bar{v}}$ & shrinkage speed\\\hline
$\rho_{+}$ & $r_{+}\bar{t}$ & catastrophe rate\\\hline
$\rho_{-}$ & $r_{-}\bar{t}$ & rescue rate\\\hline
$\rho_{n}$ & $r_{n}\bar{t}$ & nucleation rate\\\hline
$\rho_{b}$ & $r_{b}\bar{t}$ & barrier unbinding rate\\\hline
$\Lambda$ & $\frac{L}{\bar{l}}$ & distance to barrier\\\hline
\end{tabular}
\caption{Dimensionless parameters of the MT dynamical model}%
\label{tab:dparameters}%
\end{table}Note that this assignment, which has the clear advantage of
maximizing the interpretability of the non-dimensional equations, does have
the disadvantage of leaving dependencies among the parameters, as by
construction%
\begin{align}
\frac{1}{\nu_{+}}+\frac{1}{\nu_{-}}  &  =1,\label{eq:meanl}\\
\frac{\rho_{+}}{\nu_{+}}-\frac{\rho_{-}}{\nu_{-}}  &  =1.
\end{align}
To denote the \emph{independent variables} of time and length we write $\tau$
and $\lambda$ respectively. Finally, the densities, as our \emph{dependent
variables,} are denoted by $\mu_{s}\left(  \lambda,\tau|\omega_{0},\tau
_{0}\right)  =\bar{l}m_{s}\left(  \lambda\bar{l},\tau\bar{t}|\omega_{0}%
,\tau_{0}\bar{t}\right)  $.

Using these notations the steady-state results of the previous section can be
summarized as%
\begin{align}
M_{n}  &  =\frac{1}{1+\rho_{n}\left(  1-e^{-\Lambda}\right)  +\frac{\rho_{n}%
}{\rho_{b}}e^{-\Lambda}},\\
\left\langle \lambda\right\rangle _{a}  &  =\frac{\left(  1-\left(
1+\Lambda\right)  e^{-\Lambda}\right)  }{\left(  1-e^{-\Lambda}\right)  }.
\end{align}

For future reference we will also rewrite the evolution equations in a more
compact notation. To do so we treat pairs of functions $(\varphi_{+}%
(\lambda,\tau),\varphi_{-}(\lambda,\tau))$, defined on the growing and
shrinking parts of the state space respectively, as a single vector valued
function $\varphi_{s}(\lambda,\tau),s=+,-$. This allows us to write

\begin{equation}
\frac{\partial}{\partial\tau}\mu_{s}(\lambda,\tau|\omega_{0},\tau_{0}%
)=\sum_{s\prime}G_{s,s^{\prime}}^{\ast}[\mu_{s^{\prime}}(\lambda,\tau
|\omega_{0},\tau_{0})],
\end{equation}
where $G^{\ast}$ is the operator matrix
\begin{equation}
G_{s,s^{\prime}}^{\ast}=\left(
\begin{array}
[c]{cc}%
-\nu_{+}\frac{\partial}{\partial\lambda}-\rho_{+} & \rho_{-}\\
\rho_{+} & \nu_{-}\frac{\partial}{\partial\lambda}-\rho_{-}%
\end{array}
\right)  . \label{eq:Gadj}%
\end{equation}

The fact that we use the notation $G^{\ast}$, signifying the Hermitean
conjugate of the \emph{generator} $G$ of the Markov process, is conventional
when discussing the \emph{forward} Kolmogorov equation, which is the formal
term for the evolution equation for the probability densities
\cite{Williams1979}.

\section{The Mean-First Passage Time\label{sec:MFPT}}

We approach the problem of calculating the Mean-First Passage Time (MFPT) for
a microtubule to hit the boundary at a distance $L$ in three steps. We first
provide a formal solution to the problem in terms of suitably chosen set of
survival (and ruin) probabilities. We then calculate the static splitting
probabilities that describe the relative weights of the direct and indirect
paths of reaching the boundary, and finally determine the conditional MFPTs
corresponding to these sets of paths.

\subsection{Formal solution}

\label{sec:formal}We first define the \emph{survival set}, the subset of state
space excluding the boundary state, i.e. $\Omega_{\Lambda}=\Omega/B=N\cup A$.
Our goal is to determine the survival probability of the process in this set
starting at $\tau=0$ from an arbitrary active state with length $\lambda
<\Lambda$, which we denote by $S_{\Omega_{\Lambda}}\left(  \tau|\lambda
,s\right)  $. We will deconstruct this survival probability with the aid of a
number additional probabilities defined as follows: $S_{A}\left(  \tau
|\lambda,s\right)  $, the probability of surviving in the active part of the
state space, i.e. of not having passed either $\Lambda$ or shrunk back to zero
length from the initial condition, and the conditional \emph{ruin}
probabilities $R_{A}^{B}\left(  \tau|\lambda,s\right)  $ and $R_{A}^{N}\left(
\tau|\lambda,s\right)  $, being the probabilities to have exited into the
boundary state at $\lambda=\Lambda$ or the nucleation state at $\lambda=0$ at
time $\tau$ respectively, \emph{without} leaving the active state at any prior
moment. Finally, we define the survival probability to remain in the
nucleation state, which, because nucleation is a simple Poisson process, is
given by $S_{N}\left(  \tau|N\right)  =\exp\left(  -\rho_{n}\tau\right)  $.
Each of these survival or ruin probabilities has a corresponding waiting time
distribution, symbolically given by $\sigma\left(  \tau|\omega\right)
=-\frac{\partial}{\partial\tau}S\left(  \tau|\omega\right)  $ or
$\sigma\left(  \tau|\omega\right)  =\frac{\partial}{\partial\tau}R\left(
\tau|\omega\right)  $.

\begin{figure}[ptb]
\includegraphics[height=5cm]{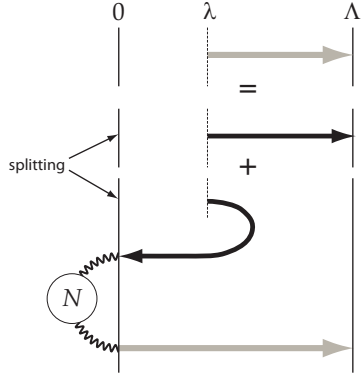}\caption{Schematic showing how the
unconditional MFPT (grey arrow) for a growing MT to reach the boundary from
starting length $\lambda$ splits into two conditional MFPTs: The first
associated with all direct paths (black arrows) from $\lambda$ to the boundary
at $\Lambda$. The second associated with all paths that shrink to zero length
without reaching the boundary, spend time in the nucleation state (wiiglly
line) and then tries again unconditionally growing from length $0$.}%
\label{fig:MFPT}%
\end{figure}

We now note the following two identities%
\begin{align}
S_{\Omega_{\Lambda}}\left(  \tau|\lambda,s\right)   &  =S_{A}\left(
\tau|\lambda,s\right)  +\int_{0}^{\tau}d\tau^{\prime}\sigma_{A}^{N}\left(
\tau^{\prime}|\lambda,s\right)  S_{\Omega_{\Lambda}}\left(  \tau-\tau^{\prime
}|N\right)  \label{eq:surva}\\
S_{\Omega_{\Lambda}}\left(  \tau|N\right)   &  =S_{N}\left(  \tau|N\right)
+\int_{0}^{\tau}d\tau^{\prime}\sigma_{N}\left(  \tau^{\prime}|N\right)
S_{\Omega_{\Lambda}}\left(  \tau-\tau^{\prime}|0,+\right)  \label{eq:survd}%
\end{align}
The first captures the fact that starting from an active state the MT survives
either by remaining active and not reaching $\Lambda$, or by shrinking back to
zero at some intermediate time and then surviving from the nucleation state.
The second states that a MT in the nucleation state survives either by
remaining in this state, or being nucleated into a growing one at an
intermediate time and then surviving from the zero-length growing state. As
survival in the active state means not exiting either at length $\Lambda$ into
the $B$ state or at length $0$ into the $N$ state we have that
\begin{equation}
S_{A}\left(  \tau|\lambda,s\right)  =1-R_{A}^{B}\left(  \tau|\lambda,s\right)
-R_{A}^{N}\left(  \tau|\lambda,s\right)
\end{equation}
As is clear from the steady state solution a MT will always leave the active
state for large enough time ($M_{n}>0$ independent of the initial conditions)
so that $S_{A}\left(  \infty|\lambda,s\right)  =0$. The ultimate conditional
ruin probabilities $R_{A}^{B}\left(  \infty|\lambda,s\right)  $ and $R_{A}%
^{N}\left(  \infty|\lambda,s\right)  $ are usually, and aptly, called
\emph{splitting probabilities}, as the total ruin probability is `split'
betwen them $R_{A}^{B}\left(  \infty|\lambda,s\right)  +R_{A}^{N}\left(
\infty|\lambda,s\right)  =1$. We can thus rewrite identity (\ref{eq:surva}) as%
\begin{multline}
S_{\Omega_{\Lambda}}\left(  \tau|\lambda,s\right)  =\left(  R_{A}^{B}\left(
\infty|\lambda,s\right)  -R_{A}^{B}\left(  \tau|\lambda,s\right)  \right)
+\left(  R_{A}^{N}\left(  \infty|\lambda,s\right)  -R_{A}^{N}\left(
\tau|\lambda,s\right)  \right)  \label{eq:survtot}\\
+\int_{0}^{\tau}d\tau^{\prime}\sigma_{A}^{N}\left(  \tau^{\prime}%
|\lambda,s\right)  S_{\Omega_{\Lambda}}\left(  \tau-\tau^{\prime}|N\right)
\end{multline}
We can now define
\begin{equation}
T_{\Omega_{\Lambda}}\left(  \lambda,s\right)  =\int_{0}^{\infty}%
d\tau\,S_{\Omega_{\Lambda}}\left(  \tau|\lambda,s\right)  =\int_{0}^{\infty
}d\tau\,\tau\,\sigma_{\Omega_{\Lambda}}\left(  \tau|\lambda,s\right)
\end{equation}
as the MFPT for the process to pass length $\Lambda$ starting from the active
state $\left(  \lambda,s\right)  $. Integrating the first two terms on the
righthand side of (\ref{eq:survtot}) over time yields
\begin{multline}
\int_{0}^{\infty}d\tau\,\left(  R_{A}^{B}\left(  \infty|\lambda,s\right)
-R_{A}^{B}\left(  \tau|\lambda,s\right)  \right)  =\int_{0}^{\infty}%
d\tau\,\tau\,\sigma_{A}^{B}\left(  \tau|\lambda,s\right)  \label{eq:Tcond}\\
=R_{A}^{B}\left(  \infty|\lambda,s\right)  \frac{\int_{0}^{\infty}d\tau
\,\tau\,\sigma_{A}^{B}\left(  \tau|\lambda,s\right)  }{\int_{0}^{\infty}%
d\tau\,\sigma_{A}^{B}\left(  \tau|\lambda,s\right)  }\equiv R_{A}^{B}\left(
\infty|\lambda,s\right)  T_{A}^{B}\left(  \lambda,s\right)
\end{multline}
which introduces the conditional MFPT of the process to exit at $\Lambda$
without ever shrinking to $0$, and similarly%
\begin{equation}
\int_{0}^{\infty}d\tau\,\left(  R_{A}^{N}\left(  \infty|\lambda,s\right)
-R_{A}^{N}\left(  \tau|\lambda,s\right)  \right)  =R_{A}^{N}\left(
\infty|\lambda,s\right)  T_{A}^{N}\left(  \lambda,s\right)
\end{equation}
Integrating the final term gives%
\begin{multline}
\int_{0}^{\infty}d\tau\,\int_{0}^{\tau}d\tau^{\prime}\sigma_{A}^{N}\left(
\tau^{\prime}|\lambda,s\right)  S_{\Omega_{\Lambda}}\left(  \tau-\tau^{\prime
}|N\right)  =\nonumber\\
\int_{0}^{\infty}d\tau^{\prime}\,\sigma_{A}^{N}\left(  \tau^{\prime}%
|\lambda,s\right)  \int_{\tau^{\prime}}^{\infty}d\tau S_{\Omega_{\Lambda}%
}\left(  \tau-\tau^{\prime}|N\right)  =R_{A}^{N}\left(  \infty|\lambda
,s\right)  T_{\Omega_{\Lambda}}\left(  N\right)
\end{multline}
The MFPT from the nucleation state is readily obtained from (\ref{eq:survd})
and yields%
\begin{equation}
T_{\Omega_{\Lambda}}\left(  N\right)  =T_{N}\left(  N\right)  +T_{\Omega
_{\Lambda}}\left(  0,+\right)  =\frac{1}{\rho_{n}}+T_{\Omega_{\Lambda}}\left(
0,+\right)
\end{equation}
where we have used that exiting the nucleation state is sure i.e.
$S_{N}\left(  \infty|N\right)  =0$. Collecting all these results then yields%
\begin{equation}
T_{\Omega_{\Lambda}}\left(  \lambda,s\right)  =R_{A}^{B}\left(  \infty
|\lambda,s\right)  T_{A}^{B}\left(  \lambda,s\right)  +R_{A}^{N}\left(
\infty|\lambda,s\right)  \left\{  T_{A}^{N}\left(  \lambda,s\right)  +\frac
{1}{\rho_{n}}+T_{\Omega_{\Lambda}}\left(  0,+\right)  \right\}
\label{eq:MFTP}%
\end{equation}
The interpretation of this result is clear: Starting from $\left(
\lambda,s\right)  $ the MT either exits directly at $\lambda=\Lambda$, which
happens with probability $R_{A}^{B}\left(  \infty|\lambda,s\right)  $ and (on
average) takes time $T_{A}^{B}\left(  \lambda,s\right)  $, or the MT first
shrinks back to $\lambda=0$ (with probability $R_{A}^{N}\left(  \infty
|\lambda,s\right)  $) which takes a time $T_{A}^{N}\left(  \lambda,s\right)
$, and then has to wait a time $\frac{1}{\rho_{n}}$ to be renucleated, after
which it takes time $T_{\Omega_{\Lambda}}\left(  0,+\right)  $ to reach
$\lambda=\Lambda$ for the first time.We illustrate this result schematically
in Figure~\ref{fig:MFPT}. 

If we now consider a MT starting at $\lambda=0$ and in the growing state $s=+$
, we can selfconsistenly solve for the MFPT\ $T_{\Omega_{\Lambda}}\left(
0,+\right)  $, which is our main result%
\begin{equation}
T_{\Omega_{\Lambda}}\left(  0,+\right)  =T_{A}^{B}\left(  0,+\right)
+\frac{R_{A}^{N}\left(  \infty|0,+\right)  }{R_{A}^{B}\left(  \infty
|0,+\right)  }\left\{  T_{A}^{N}\left(  0,+\right)  +\frac{1}{\rho_{n}%
}\right\}  . \label{eq:MFPT0p}%
\end{equation}
This latter result is in fact sufficient to solve the general problem, because
in order to reach the boundary at $\Lambda$ from length $0$ the MT first has
to pass through each intermediate length $\lambda<\Lambda$, taking time
$T_{\Omega_{\lambda}}\left(  0,+\right)  $ (note the subscript $\lambda$ to
$\Omega$ here!$)$ and then reach $\Lambda$ from there, i.e.
\begin{equation}
T_{\Omega_{\Lambda}}\left(  0,+\right)  =T_{\Omega_{\lambda}}\left(
0,+\right)  +T_{\Omega_{\Lambda}}\left(  \lambda,+\right)  ,
\end{equation}
showing that the MFPT for a MT starting in the growing state at an arbitrary
length can be expressed fully in terms of MFPTs starting from the zero-length state.

Although due to the fundamental asymmetry of the problem, there is no
corresponding simple rule for the shrinking case, the following argument shows
how we can leverage the results of the growing case to obtain a fairly compact
representation. We first introduce the survival probability of the shrinking
state with respect to rescues, which is simply given by $S_{-}\left(
\tau|-\right)  =\exp\left(  -\rho_{-}\tau\right)  $. If no rescue occurs, the
shrinking MT will hit zero length at the deterministic time $\tau_{-}\left(
\lambda\right)  =\lambda/\nu_{-}$, so that%
\begin{multline}
S_{\Omega_{\Lambda}}\left(  \tau|\lambda,-\right)  =\left(  1-H\left(
\tau-\tau_{-}\left(  \lambda\right)  \right)  \right)  \left\{  S_{-}\left(
\tau|-\right)  +\int_{0}^{\tau}d\tau^{\prime}\,\sigma_{-}\left(  \tau^{\prime
}|-\right)  S_{\Omega_{\Lambda}}\left(  \tau-\tau^{\prime}|\lambda-\nu_{-}%
\tau^{\prime},+\right)  \right\} \\
H\left(  \tau-\tau_{-}\left(  \lambda\right)  \right)  \left\{  \int_{0}%
^{\tau_{-}\left(  \lambda\right)  }d\tau^{\prime}\,\sigma_{-}\left(
\tau^{\prime}|-\right)  S_{\Omega_{\Lambda}}\left(  \tau-\tau^{\prime}%
|\lambda-\nu_{-}\tau^{\prime},+\right)  +S_{-}\left(  \tau_{-}\left(
\lambda\right)  |-\right)  S_{\Omega_{\Lambda}}\left(  \tau-\tau_{-}\left(
\lambda\right)  |N\right)  \right\}
\end{multline}
where $H\left(  x\right)  $ is the standard Heavyside function. Integrating
over all time yields the desired result%
\begin{align}
T_{\Omega_{\Lambda}}\left(  \lambda,-\right)   &  =\frac{1}{\rho_{-}}\left(
1-e^{-\rho_{-}\tau_{-}\left(  \lambda\right)  }\right)  +\rho_{-}\int
_{0}^{\tau_{-}\left(  \lambda\right)  }d\tau\,e^{-\rho_{-}\tau}T_{\Omega
_{\Lambda}}\left(  \lambda-\nu_{-}\tau,+\right)  +e^{-\rho_{-}\tau_{-}\left(
\lambda\right)  }T_{\Omega_{\Lambda}}\left(  N\right) \nonumber\\
&  =\left(  1-e^{-\rho_{-}\tau_{-}\left(  \lambda\right)  }\right)  \left\{
\frac{1}{\rho_{-}}+\frac{\int_{0}^{\tau_{-}\left(  \lambda\right)  }%
d\tau\,\sigma_{-}\left(  \tau|-\right)  T_{\Omega_{\Lambda}}\left(
\lambda-\nu_{-}\tau,+\right)  }{\int_{0}^{\tau_{-}\left(  \lambda\right)
}d\tau\,\sigma_{-}\left(  \tau|-\right)  }\right\}  +e^{-\rho_{-}\tau
_{-}\left(  \lambda\right)  }T_{\Omega_{\Lambda}}\left(  N\right) \nonumber\\
&  =R_{-}\left(  \tau_{-}\left(  \lambda\right)  |-\right)  \left\{  \frac
{1}{\rho_{-}}+\left\langle T_{\Omega_{\Lambda}}\left(  \lambda-\nu_{-}%
\tau,+\right)  \right\rangle _{\left(  \lambda,-\right)  }\right\}
+S_{-}\left(  \tau_{-}\left(  \lambda\right)  |-\right)  \left\{  \frac
{1}{\rho_{n}}+T_{\Omega_{\Lambda}}\left(  0,+\right)  \right\}
\end{align}
where $\left\langle T_{\Omega_{L}}\left(  \lambda-\nu_{-}\tau,+\right)
\right\rangle _{\left(  \lambda,-\right)  }$ is the average MFPT of an MT that
starts in a growing state after a single rescue from a shrinking state
originally at length $\lambda$ at time $0$, provided this happens before the
shrinking state hits zero length.

\subsection{The splitting probabilities}

\label{sec:split}To calculate the splitting probabilities $R_{A}^{N}\left(
\infty|\lambda,s\right)  $ and $R_{A}^{B}\left(  \infty|\lambda,s\right)  $,
we first recall from the theory of Markov processes that expectation values of
future events seen as functions of the initial time and state satisfy the
\emph{backward} Kolmogorov equation \cite{Williams1979}. Specifically, any
ruin probability $R_{K}^{C}\left(  T|\tau,\lambda,s\right)  $, where $K$ is
some subset of $\Omega$, and $C$ a conditioning event, satisfies%
\begin{equation}
\frac{\partial}{\partial\tau}R_{K}^{C}\left(  T|\tau,\lambda,s\right)
=-\frac{\partial}{\partial T}R_{K}^{C}\left(  T|\tau,\lambda,s\right)
=-\sum_{s^{\prime}}G_{s,s^{\prime}}\left[  R_{K}^{C}\left(  T|\tau
,\lambda,s^{\prime}\right)  \right]  \label{eq:back}%
\end{equation}
where the generator $G_{s,s^{\prime}}$ is the Hermitian conjugate of the
operator (\ref{eq:Gadj}), i.e.%
\begin{equation}
G_{s,s^{\prime}}=\left(
\begin{array}
[c]{cc}%
\nu_{+}\frac{\partial}{\partial\lambda}-\rho_{+} & \rho_{+}\\
\rho_{-} & -\nu_{-}\frac{\partial}{\partial\lambda}-\rho_{-}%
\end{array}
\right)  \label{eq:G}%
\end{equation}
Since our process is time homogeneous, we can of course take the initial time
to be $\tau=0$. Letting our final time $T\rightarrow\infty$, we see that
$\frac{\partial}{\partial\tau}R_{K}^{C}\left(  \infty|\tau,\lambda,s\right)
=\frac{\partial}{\partial\tau}R_{K}^{C}\left(  \infty|0,\lambda,s\right)  =0$,
so that splitting probabilities satisfy
\begin{equation}
\sum_{s^{\prime}}G_{s,s^{\prime}}\left[  R_{K}^{C}\left(  \infty
|\lambda,s^{\prime}\right)  \right]  =0,
\end{equation}
and are said to be \emph{harmonic.}

\bigskip For convenience sake we now drop the explicit mention of the final
time and set $R_{A}^{N}\left(  \lambda,s\right)  =R_{A}^{N}\left(
\infty|\lambda,s\right)  $ and $R_{A}^{B}\left(  \lambda,s\right)  =R_{A}%
^{B}\left(  \infty|\lambda,s\right)  $. We consider the latter splitting
probabilities first, and write out (\ref{eq:G}) to obtain

\begin{align}
\nu_{+}\frac{\partial}{\partial\lambda}R_{A}^{B}\left(  \lambda,+\right)
-\rho_{+}R_{A}^{B}\left(  \lambda,+\right)  +\rho_{+}R_{A}^{B}\left(
\lambda,-\right)   &  =0\\
-\nu_{-}\frac{\partial}{\partial\lambda}R_{A}^{B}\left(  \lambda,-\right)
-\rho_{-}R_{A}^{B}\left(  \lambda,-\right)  +\rho_{-}R_{A}^{B}\left(
\lambda,+\right)   &  =0
\end{align}
with the obvious boundary conditions $R_{A}^{B}\left(  \Lambda,+\right)  =1$,
$R_{A}^{B}\left(  0,-\right)  =0$. At this point it is useful to define the
mean forward and backward run-lengths%
\begin{align}
\lambda_{+}  &  =\frac{\nu_{+}}{\rho_{+}}\\
\lambda_{-}  &  =\frac{\nu_{-}}{\rho_{-}}%
\end{align}
The first equation allows us to eliminate $R_{A}^{B}\left(  \lambda,-\right)
$%
\begin{equation}
R_{A}^{B}\left(  \lambda,-\right)  =R_{A}^{B}\left(  \lambda,+\right)
-\lambda_{+}\frac{\partial}{\partial\lambda}R_{A}^{B}\left(  \lambda,+\right)
\label{eq:mprel}%
\end{equation}
Insertion into the second then yields a second order equation%
\begin{equation}
\lambda_{+}\lambda_{-}\frac{\partial^{2}}{\partial\lambda^{2}}R_{A}^{B}\left(
\lambda,+\right)  -\left(  \lambda_{-}-\lambda_{+}\right)  \frac{\partial
}{\partial\lambda}R_{A}^{B}\left(  \lambda,+\right)  =0
\end{equation}
or equivalently%
\begin{equation}
\frac{\partial^{2}}{\partial\lambda^{2}}R_{A}^{B}\left(  \lambda,+\right)
-\frac{\partial}{\partial\lambda}R_{A}^{B}\left(  \lambda,+\right)  =0
\end{equation}
as (\ref{eq:meanl}) implies
\begin{equation}
\frac{1}{\lambda_{+}}-\frac{1}{\lambda_{-}}=1
\end{equation}
We then obtain the following solutions%
\begin{align}
R_{A}^{B}\left(  \lambda,+\right)   &  =\frac{e^{\lambda}-\left(
1-\lambda_{+}\right)  }{e^{\Lambda}-\left(  1-\lambda_{+}\right)  }\\
R_{A}^{B}\left(  \lambda,-\right)   &  =\frac{\left(  e^{\lambda}-1\right)
\left(  1-\lambda_{+}\right)  }{e^{\Lambda}-\left(  1-\lambda_{+}\right)  }%
\end{align}

In a fully similar manner the corresponding quantities $R_{A}^{N}\left(
\lambda,s\right)  $ are also readily determined%
\begin{align}
R_{A}^{N}\left(  \lambda,+\right)   &  =\frac{e^{\Lambda}-e^{\lambda}%
}{e^{\Lambda}-\left(  1-\lambda_{+}\right)  }\\
R_{A}^{N}\left(  \lambda,-\right)   &  =\frac{e^{\Lambda}-\left(
1-\lambda_{+}\right)  e^{\lambda}}{e^{\Lambda}-\left(  1-\lambda_{+}\right)  }%
\end{align}
One checks that these forms satisfy the a priori requirements $R_{A}%
^{B}\left(  \lambda,+\right)  +R_{A}^{N}\left(  \lambda,+\right)  =R_{A}%
^{B}\left(  \lambda,-\right)  +R_{A}^{N}\left(  \lambda,-\right)  =1$, which
follow from the fact that the ultimate ruin of an MT on a finite length
interval is sure. These splitting probabilities were also derived in
\cite{Holy1997} by considering Laplace transforms of recurrence relations
satisfied by the probability density.

\subsection{The conditional MFPTs}

\label{sec:CMFPT}With the splitting probabilities determined, we can directly
calculate the conditional MFPTs by solving a time integrated form of the
backward equation. Indeed, integrating (\ref{eq:back}) over the final time $T$
, and recalling that $R_{K}^{C}\left(  \tau|\tau,\lambda,s\right)  =0$, yields
as a first step%
\begin{equation}
-\int_{\tau}^{\infty}dT\frac{\partial}{\partial T}R_{K}^{C}\left(
T|\tau,\lambda,s\right)  =-R_{K}^{C}\left(  \infty|\lambda,s\right)
=-\int_{\tau}^{\infty}dT\sum_{s^{\prime}}G_{s,s^{\prime}}\left[  \,R_{K}%
^{C}\left(  T|\tau,\lambda,s^{\prime}\right)  \right]  . \label{eq:BackR}%
\end{equation}
At this point we would like to interchange the integration and the operation
of $G_{s,s^{\prime}}$, but as $R_{K}^{C}\left(  T|\tau,\lambda,s^{\prime
}\right)  $ tends to a constant for $T\rightarrow\infty$ this is not directly
possible. However, we can use the fact that the splitting probabilities are
harmonic, i.e $G_{s,s^{\prime}}\left[  R_{K}^{C}\left(  \infty|\lambda
,s^{\prime}\right)  \right]  =0$, and the linearity $G_{s,s^{\prime}}$ to
obtain the identity
\begin{equation}
-G_{s,s^{\prime}}\left[  \,R_{K}^{C}\left(  T|\tau,\lambda,s^{\prime}\right)
\right]  =G_{s,s^{\prime}}\left[  R_{K}^{C}\left(  \infty|\lambda,s^{\prime
}\right)  -R_{K}^{C}\left(  T|\tau,\lambda,s^{\prime}\right)  \right]  .
\end{equation}
Substitution of this identity into Eq.~(\ref{eq:BackR}) yields an integrable
argument exactly of the form previously encountered in Eq.~(\ref{eq:Tcond}),
so that%
\begin{multline}
-\int_{\tau}^{\infty}dT\sum_{s^{\prime}}G_{s,s^{\prime}}\left[  \,R_{K}%
^{C}\left(  T|\tau,\lambda,s^{\prime}\right)  \right]  =\int_{\tau}^{\infty
}dT\sum_{s^{\prime}}G_{s,s^{\prime}}\left[  R_{K}^{C}\left(  \infty
|\lambda,s^{\prime}\right)  -R_{K}^{C}\left(  T|\tau,\lambda,s^{\prime
}\right)  \right] \\
=\sum_{s^{\prime}}G_{s,s^{\prime}}\left[  \int_{\tau}^{\infty}dT\left\{
R_{K}^{C}\left(  \infty|\lambda,s^{\prime}\right)  -R_{K}^{C}\left(
T|\tau,\lambda,s^{\prime}\right)  \right\}  \right]  =\sum_{s^{\prime}%
}G_{s,s^{\prime}}\left[  R_{K}^{C}\left(  \infty|\lambda,s^{\prime}\right)
T_{K}^{C}\left(  \lambda,s^{\prime}\right)  \right]  \label{eq:BackR2}%
\end{multline}
Combining, Eqs.~(\ref{eq:BackR}) and (\ref{eq:BackR2}), yields the sought
after relation%
\begin{equation}
\sum_{s^{\prime}}G_{s,s^{\prime}}\left[  R_{K}^{C}\left(  \infty
|\lambda,s^{\prime}\right)  T_{K}^{C}\left(  \lambda,s^{\prime}\right)
\right]  =-R_{K}^{C}\left(  \infty|\lambda,s\right)  , \label{eq:TcondG}%
\end{equation}
which together with appropriate boundary conditions yields a closed form
equation for the conditional MFPTs $T_{K}^{C}\left(  \lambda,s^{\prime
}\right)  $.

We now apply Eq.~(\ref{eq:TcondG}) to our problem, starting with the case of
exiting at $\Lambda$ we have%

\begin{align}
\lambda_{+}\frac{\partial}{\partial\lambda}R_{A}^{B}\left(  \lambda,+\right)
T_{A}^{B}\left(  \lambda,+\right)  -R_{A}^{B}\left(  \lambda,+\right)
T_{A}^{B}\left(  \lambda,+\right)  +R_{A}^{B}\left(  \lambda,-\right)
T_{A}^{B}\left(  \lambda,-\right)   &  =-\frac{1}{\rho_{+}}R_{A}^{B}\left(
\lambda,+\right) \\
-\lambda_{-}\frac{\partial}{\partial\lambda}R_{A}^{B}\left(  \lambda,-\right)
T_{A}^{B}\left(  \lambda,-\right)  -R_{A}^{B}\left(  \lambda,-\right)
T_{A}^{B}\left(  \lambda,-\right)  +R_{A}^{B}\left(  \lambda,+\right)
T_{A}^{B}\left(  \lambda,+\right)   &  =-\frac{1}{\rho_{-}}R_{A}^{B}\left(
\lambda,-\right)
\end{align}
with boundary conditions $R_{A}^{B}\left(  \Lambda,+\right)  T_{A}^{B}\left(
\Lambda,+\right)  =R_{A}^{B}\left(  0,-\right)  T_{A}^{B}\left(  0,-\right)
=0$. Eliminating $R_{A}^{B}\left(  \lambda,-\right)  T_{A}^{B}\left(
\lambda,-\right)  $ and introducing the shorthand $\Theta_{A}^{B}\left(
\lambda,+\right)  =R_{A}^{B}\left(  \lambda,+\right)  T_{A}^{B}\left(
\lambda,+\right)  $ we find the following inhomogeneous second order equation%
\begin{multline}
\frac{\partial^{2}}{\partial\lambda^{2}}\Theta_{A}^{B}\left(  \lambda
,+\right)  -\frac{\partial}{\partial\lambda}\Theta_{A}^{B}\left(
\lambda,+\right)  =\label{eq:CMFPTp}\\
-\frac{1}{\lambda_{+}\rho_{+}}\frac{\partial}{\partial\lambda}R_{A}^{B}\left(
\lambda,+\right)  -\frac{1}{\lambda_{+}\lambda_{-}}\left(  \frac{1}{\rho_{+}%
}R_{A}^{B}\left(  \lambda,+\right)  +\frac{1}{\rho_{-}}R_{A}^{B}\left(
\lambda,-\right)  \right)  \equiv A_{A}^{B}\left(  \lambda,+\right)
\end{multline}
Using Eq.~(\ref{eq:mprel}) we can eliminate $R_{A}^{B}\left(  \lambda
,-\right)  $ in the inhomogeneous term, so that%
\begin{equation}
A_{A}^{B}\left(  \lambda,+\right)  =\left(  \frac{1}{\lambda_{-}\rho_{-}%
}-\frac{1}{\lambda_{+}\rho_{+}}\right)  \frac{\partial}{\partial\lambda}%
R_{A}^{B}\left(  \lambda,+\right)  -\frac{1}{\lambda_{+}\lambda_{-}}\left(
\frac{1}{\rho_{+}}+\frac{1}{\rho_{-}}\right)  R_{A}^{B}\left(  \lambda
,+\right)
\end{equation}
The boundary equations for the resulting equation are%
\begin{align}
\Theta_{A}^{B}\left(  \Lambda,+\right)   &  =0\label{eq:CMFPTbc1}\\
\lambda_{+}\frac{\partial}{\partial\lambda}\Theta_{A}^{B}\left(  0,+\right)
-\Theta_{A}^{B}\left(  0,+\right)   &  =-\frac{1}{\rho_{+}}R_{A}^{B}\left(
0,+\right)  \label{eq:CMFPTbc2}%
\end{align}
The equation for the conditional MFPT for exiting at $0$ is the formally the
same as (\ref{eq:CMFPTp}), together with boundary conditions
(\ref{eq:CMFPTbc1}) and (\ref{eq:CMFPTbc2}), but with $\Theta_{A}^{N}\left(
\lambda,+\right)  $ and $R_{A}^{N}\left(  \lambda,+\right)  $ substituted for
$\Theta_{A}^{B}\left(  \lambda,+\right)  $ and $R_{A}^{B}\left(
\lambda,+\right)  $ respectively. These inhomogeneous linear second order
equations are readily solved, and we present the resultant rather unwieldy
expressions in the appendix. As an explicit check we consider the probability
of a growing MT to shrink back to the origin, in the absence of the boundary
\begin{equation}
\lim_{\Lambda\rightarrow\infty}\Theta_{A}^{N}\left(  \lambda,+\right)
=\lim_{\Lambda\rightarrow\infty}T_{A}^{N}\left(  \lambda,+\right)  \equiv
T\left(  \lambda,+\right)  =1+\lambda\frac{\left(  \rho_{+}+\rho_{-}\right)
}{\left(  \rho_{+}\nu_{-}-\rho_{-}\nu_{+}\right)  }%
\end{equation}
Upon redimensionalizing, this expression is identical to the one derived
earlier by Bicout from the full time and space-dependent survival probability
\cite{Bicout1997}. As an aside, we note that the result $T(0,+)=1$ shows that
the time-scale we have adopted is indeed that of the origin return time of an
unconstrained MT, as already stated in Section \ref{sec:SS}.

\subsection{Application to biological data}

In order to get a feel what the results derived above mean in real-world
terms, we apply them to two sets of fairly well characterized kinetic
parameters for MTs, one derived from observations on fission yeast
\cite{Janson2007}, and one on interphase Tobacco Bright Yellow-2 plant culture
cells \cite{Vos2004}. These data sets are summarized in Table
\ref{tab:bparameters}. \begin{table}[ptb]%
\begin{tabular}
[c]{|l|c|c|}\hline
Parameter & Yeast & Plant\\\hline\hline
Growth speed( $\mu$m/min) & $2.4$ & $4.8$\\\hline
Shrinking speed ($\mu$m/min) & $9.6$ & $9.6$\\\hline
Catastrophe rate (/min) & $0.3$ & $0.28$\\\hline
Rescue rate (/min) & $-$ & $0.42$\\\hline
Nucleation rate (/min) & $0.15$ & $0.15^{\dagger}$\\\hline
Mean length (no boundary) ( $\mu$m) & $8$ & $68.5$\\\hline
Expected lifetime (no boundary) (min) & $4.17$ & $21.43$\\\hline
\end{tabular}
\caption{Table of MT dynamical parameters for yeast and interphase plant
cells. {\footnotesize {$\dagger$: In the absence of available data we take
this number equal to that of yeast.} }}%
\label{tab:bparameters}%
\end{table}

\begin{figure}[ptb]
\includegraphics[width=\textwidth]{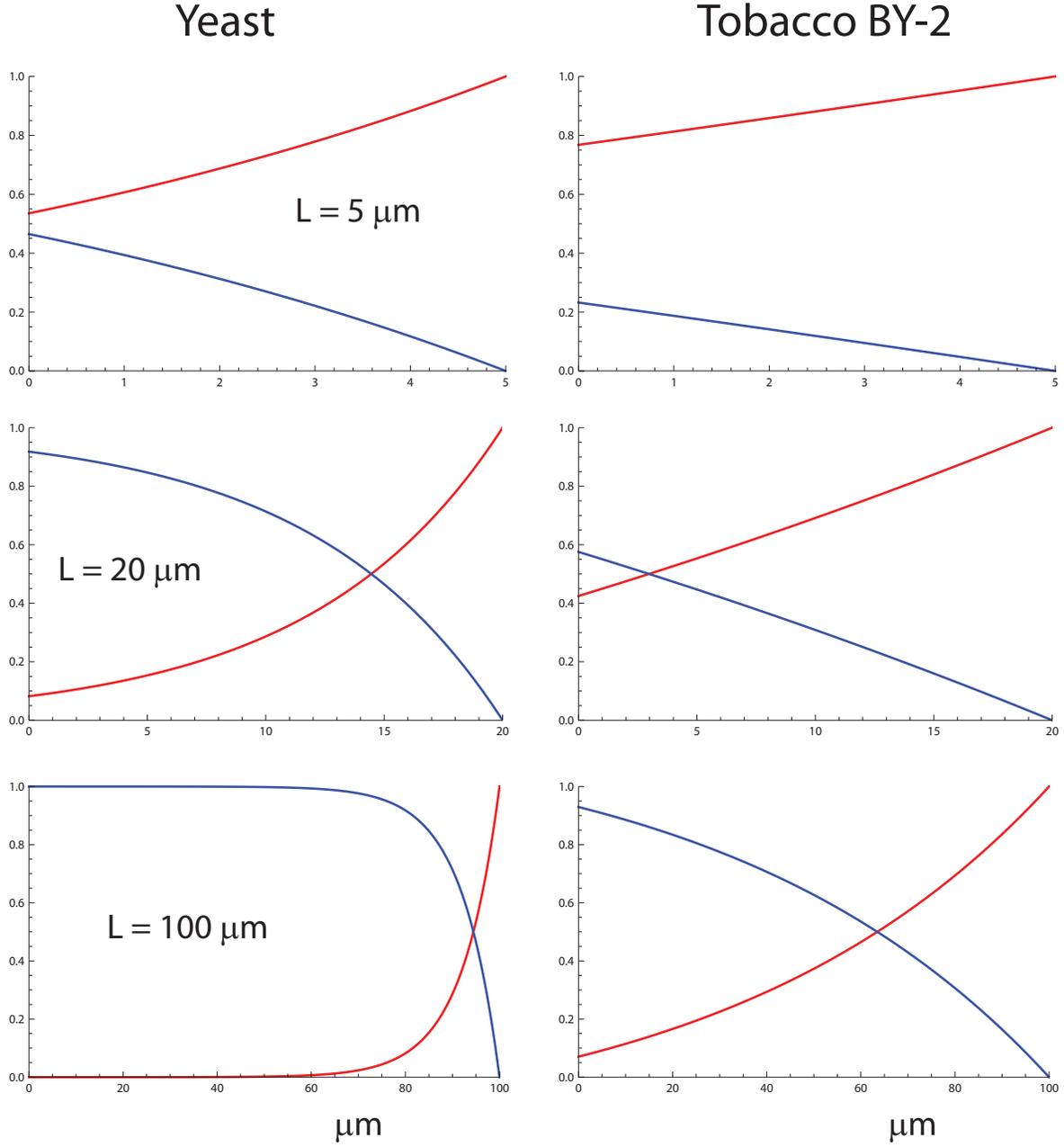}\caption{Splitting
probabilities $R_{A}^{B}(l,+)$ (red line) and $R_{A}^{N}(l,+)$ (blue line) as
a function of the initial length $l$ for different values of the distance
$L$.}%
\label{fig:split}%
\end{figure}

We now confront these two types of MTs with boundaries located at $5\mu m$,
smaller than both mean lengths in the absence of boundaries and comparable to
half the length of a fission yeast cell, $20\mu m$ double the mean length for
the yeast MT and still significantly smaller than that of the Tobacco BY-2
MTs, and $100\mu m$ of the order of typical lengthscale of a Tobacco BY-2
cell. We first consider the splitting probabilities $R_{A}^{B}\left(
l,+\right)  $ and $R_{A}^{N}\left(  l,+\right)  $,which we plot in Figure
\ref{fig:split}. We see that for the smallest boundary distance $L=5\mu m$ both for
yeast and plant MTs the probability to reach the boundary from zero length is
already appreciable, and increases roughly linearly with starting length,
consistent with it being dominated by uninterrupted growth. As we increase the
boundary distance, the probabilties depend more strongly non-linearly on the
starting length. This is most striking for the case of yeast at $L=100\mu m$,
where $R_{A}^{B}\left(  l,+\right)  $ is essentially $0$ until the starting
length is within the natural length $\bar{l}\approx8\mu m$ from the boundary.

\begin{figure}[ptb]
\includegraphics[width=\textwidth]{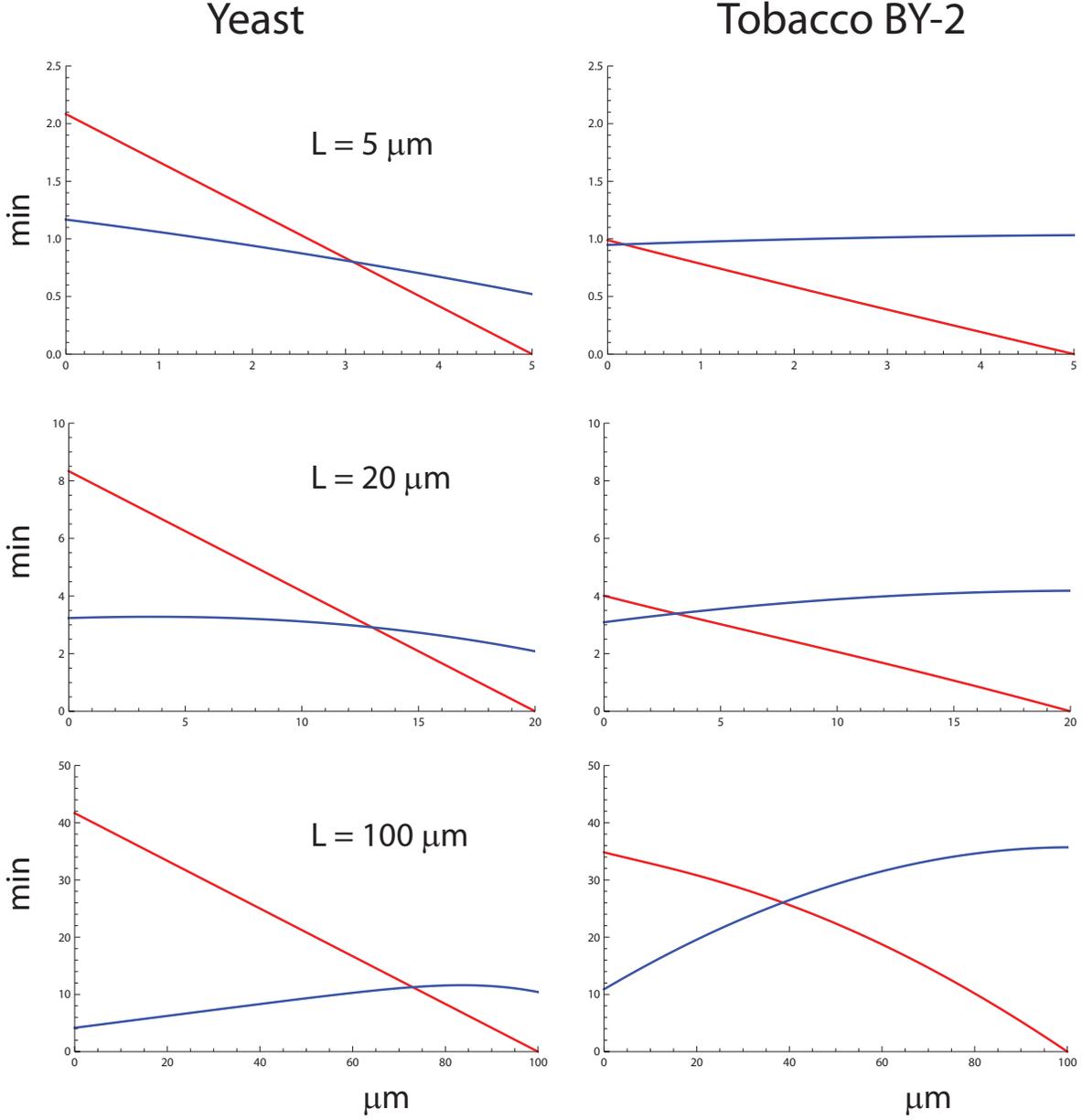}\caption{Conditional MFPTs
$T_{A}^{B}\left(  l,+\right)  $ (red line) and $T_{A}^{N}\left(  l,+\right)  $
(blue line) as a function of the initial length $l$ for different values of
the distance $L$.}%
\label{fig:cMFPT}%
\end{figure}

Next, we turn to the conditional MFPTs $T_{A}^{B}\left(  l,+\right)  $ and
$T_{A}^{N}\left(  l,+\right)  $. Here, we first need to take \ a little care,
as for yeast\ the rescue probability vanishes ($r_{-}=0)$, so that the
backward runlength $l_{-}$ is ill-defined. One can of course go through the
procedure in Section \ref{sec:CMFPT} again setting $\rho_{-}=0$ at the outset
in Eq.~(\ref{eq:G}). However, in this case the conditional MFPTs are also
readily determined from first principles%
\begin{align}
r_{-} &  =0:T_{A}^{B}\left(  l,+\right)  =\frac{\left(  L-l\right)  }{v_{+}%
}\label{eq:cMFTPYa}\\
r_{-} &  =0:T_{A}^{N}\left(  l,+\right)  =\frac{\int_{0}^{\frac{\left(
L-l\right)  }{v_{+}}}dt\,e^{-r_{+}t}\left\{  t+\frac{\left(  l+v_{+}t\right)
}{v_{-}}\right\}  }{\int_{0}^{\frac{\left(  L-l\right)  }{v_{+}}}%
dt\,e^{-r_{+}t}}=\frac{l}{v_{-}}+\bar{t}\frac{\left(  1-e^{-\frac{L-l}{\bar
{l}}}\left(  1+\frac{L-l}{\bar{l}}\right)  \right)  }{1-e^{-\frac{L-l}{\bar
{l}}}}\label{eq:cMFTPYb}%
\end{align}
where (\ref{eq:cMFTPYa}) follows because a non-rescuable MT can only reach the
boundary without first shrinking away by growing towards it deterministically,
and (\ref{eq:cMFTPYb}) is obtained by averaging (i) the time to experience a
catastrophe before reaching the boundary plus (ii) the time to shrink to zero
length from that moment on over the ensemble of histories that do not reach
the boundary. These two approaches indeed give the same results, which serves
as another independent check on the general formalism. Figure \ref{fig:cMFPT}
shows the resulting passage times. A perhaps at first sight puzzling feature of these results is the
decrease of $T_{A}^{N}\left(  l,+\right)  $ for increasing starting length
$l$, which is evident for the yeast case. This, however, is a direct
consequence of the conditioning on shrinking back without reaching the
boundary: If the starting length is within the forward runlength $l_{+}$ from
the boundary, a conditioned MT must rapidly undergo a catastrophe after which
it deterministically shrinks back to zero length. The conditioned return time
(the second term on the far right hand side of Eq.~(\ref{eq:cMFTPYb})) is
therefore a strongly non-linearly decreasing function of the distance to the
boundary, whereas the time to deterministically shrink back from the starting
length $l/v_{-}$ only increases linearly with length.

Finally, in Table \ref{tab:MFPT} we give the MFPT $T_{\Omega_{L}}\left(
0,+\right) $ for reaching the barrier starting from zero length in the growing
state as calculated from Eq.~(\ref{eq:MFPT0p}).

\begin{table}[ptb]%
\begin{tabular}
[c]{|l|l|l|}\hline
$L$ & Yeast & Plant\\\hline\hline
$5\mu m$ & $8.89$ & $3.29$\\\hline
$20\mu m$ & $119.1$ & $17.23$\\\hline
$100\mu m$ & $2.9\times10^{6}$ & $266.55$\\\hline
\end{tabular}
\caption{Table of the MFPT $T_{\Omega_{L}}(0,+)$ (in min) as a function of the
distance $L$.}%
\label{tab:MFPT}%
\end{table}We see that for a yeast MT the largest boundary distance $L=100\mu
m$ is effectively unbridgeable, and that even the plant MT needs on average
$\approx10$ times its natural lifetime $\bar{t}\approx20\min$ to first reach
the boundary, although it is only $\approx1.5$ times its natural length of
$\bar{l}\approx65\mu m$.

\section{The Gopalakrishnan-Govindan search-and-capture model}

\label{sec:GG}The first-passage-time model Gopalakrishnan and Govindan
recently introduced (\cite{Gopalakrishnan2011}, hereafter referred to as GG)
considers the problem of the `capture' of a chromosome by a MT 'searching' for
it. It has the following ingredients: The MT is nucleated at a rate $r_{n}$
from a centrosome in an arbitrary direction into a cone with solid angle
opening of $\Delta\Omega$. The centrosome is located at a distance $d$ from
the chromosome, which has a cross-sectional area $a$ and therefore subtends a
solid angle $\Delta\Omega_{c}=a/d^{2}$ as seen from the centrosome. The
probability of being nucleated into a direction in which the target can
possibly be hit is therefore given by $p_{c}=\Delta\Omega_{c}/\Delta\Omega$.
When the MT is nucleated outside of the `\emph{target cone}', it can
potentially grow until it hits a cell boundary located at a distance we will
call $D$ from the centrosome. At this boundary, the MT is initially stalled,
but experiences an increased catastrophe rate $r_{b}>r_{+}$.

\begin{figure}[ptb]
\includegraphics[width=\textwidth]{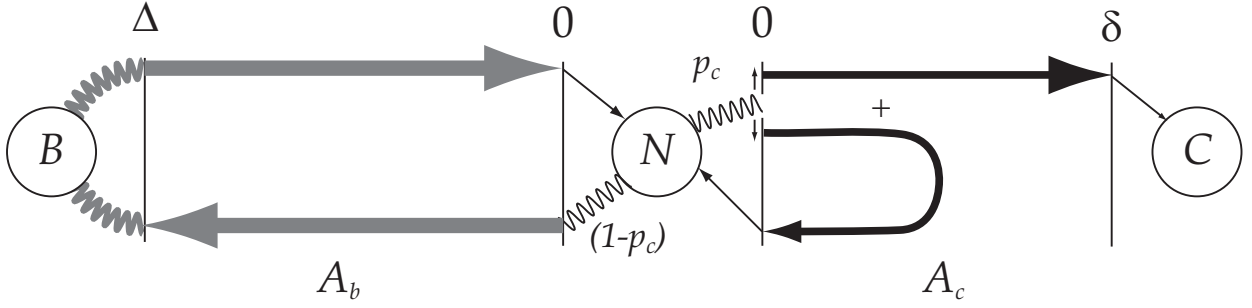}\caption{Schematic illustration of the
structure of the search process in the Gopalakrishnan-Govindan
search-and-capture model. From the nucleation state the MT must either perform
(with probability $1-p_{c}$) a fruitless search (gray arrows) in the
directions where it can interact with the cell boundary $A_{b}$, or (with
probability $p_{c}$) it is nucleated in the right direction and then either
directly traverses $A_{c}$ to reach the chromosome $C$ or shrinks back to zero
length without reaching the target and must try again.}%
\label{fig:GG}%
\end{figure}

We will now revisit this model, using the formalism derived in the previous
sections. The state space of this model is conveniently represented by
$\Omega=N\cup A_{b}\cup B\cup A_{c}\cup C$. Here, as before, $N$ is the
nucleation state, $A_{b}$ are the active states with lengths in the interval
$[0,D]$ in the directions that do not `see' the target, $B$ the state of being
at the cell boundary, $A_{c}$ the active states with lengths in the interval
$[0,d]$ and directions within the target cone, and, finally, $C$ the state of
being on the target chromosome. We non-dimensionalize using the same
prescription as in Section \ref{sec:dimen}, denoting the additional parameters
needed by $\Delta=D/\bar{l}$ and $\delta=d/\bar{l}$. Using the results of
Section~\ref{sec:formal} we can immediately write down an expression for the
search-time starting from the nucleation state%
\begin{multline}
T_{\Omega/C}\left(  N\right)  =\frac{1}{\rho_{n}}+\left(  1-p_{c}\right)
\left\{  T_{A_{b}\cup B}\left(  \left(  0,+\right)  _{A_{b}}\right)
+T_{\Omega/C}\left(  N\right)  \right\} \label{eq:Tc}\\
+p_{c}\left\{  R_{A_{c}}^{C}\left(  \left(  0,+\right)  _{A_{c}}\right)
T_{A_{c}}^{C}\left(  \left(  0,+\right)  _{A_{c}}\right)  +R_{A_{c}}%
^{N}\left(  \left(  0,+\right)  _{A_{c}}\right)  \left\{  T_{A_{c}}^{N}\left(
\left(  0,+\right)  _{A_{c}}\right)  +T_{\Omega/C}\left(  N\right)  \right\}
\right\}
\end{multline}
The logic of this equation is simple. Starting from the nucleation state the
MT (on average) waits $1/\rho_{n}$ before being nucleation. With probability
$\left(  1-p_{c}\right)  $ the nucleation will be in a direction that can not
hit the target. In that case the MT will spend the origin-return time
$T_{A_{b}\cup B}\left(  \left(  0,+\right)  _{A_{b}}\right)  $ in this part of
state space before shrinking back to zero length and starting again from the
nucleation state. With probability $p_{c}$ the initial nucleation is inside
the target cone. In that case the MT either hits the target, without first
shrinking back to zero-length, with probability $R_{A_{c}}^{C}\left(  \left(
0,+\right)  _{A_{c}}\right)  $ taking time $T_{A_{c}}^{C}\left(  \left(
0,+\right)  _{A_{c}}\right)  $, or, with probability $R_{A_{c}}^{N}\left(
\left(  0,+\right)  _{A_{c}}\right)  =1-R_{A_{c}}^{C}\left(  \left(
0,+\right)  _{A_{c}}\right)  $ shrinking back to zero-length before hitting
the target, taking time $T_{A_{c}}^{N}\left(  \left(  0,+\right)  _{A_{c}%
}\right)  $ and then trying again from the nucleation state. This process is
illustrated in Figure~\ref{fig:GG}. 

Extracting $T_{\Omega/C}\left(  N\right)  $ from the relation (\ref{eq:Tc})
yields%
\begin{multline}
p_{c}R_{A_{c}}^{C}\left(  \left(  0,+\right)  _{A_{c}}\right)  T_{\Omega
/C}\left(  N\right)  =\frac{1}{\rho_{+}}+\left(  1-p_{c}\right)  T_{A_{b}\cup
B}\left(  \left(  0,+\right)  _{A_{b}}\right) \\
+p_{c}\left\{  R_{A_{c}}^{C}\left(  \left(  0,+\right)  _{A_{c}}\right)
T_{A_{c}}^{C}\left(  \left(  0,+\right)  _{A_{c}}\right)  +R_{A_{c}}%
^{N}\left(  \left(  0,+\right)  _{A_{c}}\right)  T_{A_{c}}^{N}\left(  \left(
0,+\right)  _{A_{c}}\right)  \right\}  \label{eq:Tsearch}%
\end{multline}
As we show in Appendix \ref{sec:GGapp}, this is, apart from the changed
notation, precisely the result derived by GG (their Eq.~(34) in Appendix A)
from an explicit sum-over-histories argument. We also note the calculation of
the fruitless search time, $T_{A_{b}\cup B}\left(  \left(  0,+\right)
_{A_{b}}\right)  $, in the directions not containing the target, is also
readily simplified using the methods presented here%
\begin{multline}
T_{A_{b}\cup B}\left(  \left(  0,+\right)  _{A_{b}}\right)  =R_{A_{b}}%
^{N}\left(  \left(  0,+\right)  _{A_{b}}\right)  T_{A_{b}}^{N}\left(  \left(
0,+\right)  _{A_{b}}\right) \\
+R_{A_{b}}^{B}\left(  \left(  0,+\right)  _{A_{b}}\right)  \left\{  T_{A_{b}%
}^{B}\left(  \left(  0,+\right)  _{A_{b}}\right)  +\frac{1}{\rho_{b}}%
+T_{A_{b}\cup B}\left(  \left(  \Delta,-\right)  _{A_{b}}\right)  \right\}  ,
\end{multline}
where in turn%
\begin{align}
T_{A_{b}\cup B}\left(  \left(  \Delta,-\right)  _{A_{b}}\right)   &
=R_{A_{b}}^{N}\left(  \left(  \Delta,-\right)  _{A_{b}}\right)  T_{A_{b}}%
^{N}\left(  \left(  \Delta,-\right)  _{A_{b}}\right) \\
&  +R_{A_{b}}^{B}\left(  \left(  \Delta,-\right)  _{A_{b}}\right)  \left\{
T_{A_{b}}^{B}\left(  \left(  \Delta,-\right)  _{A_{b}}\right)  +\frac{1}%
{\rho_{b}}+T_{A_{b}\cup B}\left(  \left(  \Delta,-\right)  _{A_{b}}\right)
\right\}  .\nonumber
\end{align}
The latter expression allows $T_{A_{b}\cup B}\left(  \left(  \Delta,-\right)
_{A_{b}}\right)  $ to be expressed solely of splitting probabilities and
conditional MFPTs. While GG use an ingenious symmetry argument interpreting a
shrinking MT as a growing `anti'-MT to calculate these latter quantities, we
point out that they can also be obtained in a straightforward manner from the
differential equations presented in Sections \ref{sec:split} and
\ref{sec:CMFPT}.

\section{Conclusions and outlook}

Our aim was to present a structured approach to the problem of MTs interacting
with boundaries. To this end we relied exclusively on `backward' techniques,
focussing on survival probabilities and their associated boundary value
problems. The upshot of this approach is that it allows one to decompose a
complex MFPT problem \emph{a priori} into closed form self-consistency problem
involving a small set of relevant splitting probabilities and conditional
MFPTs that readily follow from a proper disjoint decomposition of the state
space. The utility of this approach is illustrated by its application to the
Gopala\-krishnan-\-Govindan model, where the key decomposition of the search
time in terms of the time spent fruitlessly searching in the wrong directions,
waiting in the nucleation state and finally reaching the target is the
starting point of the calculation, rather than, as in
\cite{Gopalakrishnan2011}, the result of collecting the results of
intermediate steps in the calculation. We hope that the technique presented
will serve as a convenient starting point for future applications to current
problems in microtubule cytoskeleton organization.

\acknowledgments This work is part of the research program of the
\textquotedblleft Stichting voor Fundamenteel Onderzoek der Materie
(FOM)\textquotedblright, which is financially supported by the
\textquotedblleft Nederlandse Organisatie voor Wetenschappelijk Onderzoek
(NWO)\textquotedblright.

\appendix

\newpage

\section{Explicit solutions of the conditional MFPTs}

In order to give a fairly concise expression for the conditional MFPTs we need
to introduce a number of convenient coefficients%

\begin{align}
\alpha &  =\lambda_{+}-1\\
\beta &  =\frac{1}{\lambda_{+}\lambda_{-}}\left(  \frac{1}{\rho_{+}}+\frac
{1}{\rho_{-}}\right) \\
\gamma &  =\frac{\lambda_{+}}{\rho_{-}}-\frac{\lambda_{-}}{\rho_{+}}\\
\delta &  =\lambda_{+}\left\{  \beta\left(  1-\alpha\right)  +\gamma\right\}
\\
\varepsilon &  =1+\alpha-\beta\rho_{+}+\alpha^{2}\beta\rho_{+}-\gamma\rho
_{+}-\alpha\gamma\rho_{+}\nonumber\\
&  =\lambda_{+}-\rho_{+}\delta
\end{align}
With these definitions we find%

\begin{align}
T_{A}^{N}\left(  \lambda,+\right)   &  =\beta\lambda_{+}+\beta\lambda+\\
&  \frac{1}{e^{\Lambda}-e^{\lambda}}\left\{  -e^{\lambda}\left(  2\beta
+\gamma\right)  \left(  \Lambda-\lambda\right)  +R_{A}^{N}\left(
\lambda,+\right)  \left(  \frac{1}{\rho_{+}}\left(  e^{\Lambda}-1\right)
+\delta-\alpha\left(  2\beta+\gamma\right)  \Lambda\right)  \right\}
\nonumber\\
T_{A}^{B}\left(  \lambda,+\right)   &  =\frac{1}{\rho_{+}}\frac{1}{e^{\lambda
}+\alpha}\left\{
\begin{array}
[c]{c}%
\varepsilon+\alpha\rho_{+}\left(  \beta+\gamma\right)  \Lambda+\alpha\beta
\rho_{+}\lambda+\left(  \beta+\gamma\right)  \rho_{+}\left(  \Lambda
-\lambda\right)  e^{\lambda}\\
-R_{A}^{B}\left(  \lambda,+\right)  \left(  \varepsilon+\alpha\rho_{+}\left(
\beta+\gamma\right)  \Lambda+\alpha\beta\rho_{+}\Lambda\right)
\end{array}
\right\}
\end{align}
In some cases we can use the known relations%

\begin{align}
1  &  =\frac{1}{\lambda_{+}}-\frac{1}{\lambda_{-}}\\
1  &  =\frac{1}{\lambda_{+}\rho_{+}}+\frac{1}{\lambda_{-}\rho_{-}}%
\end{align}
to simplify even further. An example is the observation that%
\begin{equation}
\beta\lambda_{+}=\frac{1}{\lambda_{-}}\left(  \frac{1}{\rho_{+}}+\frac{1}%
{\rho_{-}}\right)  =\left(  \frac{1}{\lambda_{+}}-1\right)  \frac{1}{\rho_{+}%
}+\frac{1}{\lambda_{-}\rho_{-}}=1-\frac{1}{\rho_{+}}%
\end{equation}
Taking the limit $\Lambda\rightarrow\infty$ then yields, as in this limit
$R_{A}^{N}\left(  \lambda,+\right)  =1$,
\begin{equation}
T_{A}^{N}\left(  0,+\right)  =1
\end{equation}
as claimed in the main text.

\section{Formal correspondence with the Gopala\-krishnan-\- Govindan model}

\label{sec:GGapp}Here we provide the translation between the results of GG and
our own, by noting the following correspondences.
\begin{equation}%
\begin{tabular}
[c]{|c|c|p{8cm}|}\hline
GG & This work & Description\\\hline\hline
$\varPhi\left(  d,T\right)  $ & $\sigma_{A_{c}}^{C}\left(  T|\left(
0,+\right)  _{A_{c}}\right)  $ & Conditional waiting time distribution
reaching target without shrinking back to zero\\\hline
$Q_{d}\left(  T\right)  $ & $\sigma_{A_{c}}^{N}\left(  T|\left(  0,+\right)
_{A_{c}}\right)  $ & Conditional waiting time distribution shrinking back to
zero without reaching target\\\hline
$Q_{R}\left(  T\right)  $ & $\sigma_{A_{b}}^{N}\left(  T|\left(  0,+\right)
_{A_{b}}\right)  $ & Conditional waiting time distribution shrinking back to
zero without reaching boundary\\\hline
$\Psi\left(  T\right)  $ & $\sigma_{A_{b}\cup B}\left(  T|\left(  0,+\right)
_{A_{b}},\exists\tau<T:\lambda_{\tau}=\Delta\right)  $ & Waiting time
distribution return to length zero after reaching the boundary at least
once\\\hline
\end{tabular}
\end{equation}
From these correspondences we derive the identities%
\begin{align}
\tilde{\varPhi}\left(  d,0\right)   &  =\int_{0}^{\infty}dT\sigma_{A_{c}}%
^{C}\left(  T|\left(  0,+\right)  _{A_{c}}\right)  =R_{A_{c}}^{C}\left(
\left(  0,+\right)  _{A_{c}}\right) \\
\tilde{\varPhi}^{\prime}\left(  d,0\right)   &  =-\int_{0}^{\infty
}dT\,T\,\sigma_{A_{c}}^{C}\left(  T|\left(  0,+\right)  _{A_{c}}\right)
=-R_{A_{c}}^{C}\left(  \left(  0,+\right)  _{A_{c}}\right)  T_{A_{c}}%
^{C}\left(  \left(  0,+\right)  _{A_{c}}\right) \\
\tilde{Q}^{\prime}\left(  d,0\right)   &  =-\int_{0}^{\infty}dT\,T\,\sigma
_{A_{c}}^{N}\left(  T|\left(  0,+\right)  _{A_{c}}\right)  =-R_{A_{c}}%
^{N}\left(  \left(  0,+\right)  _{A_{c}}\right)  T_{A_{c}}^{N}\left(  \left(
0,+\right)  _{A_{c}}\right) \\
\tilde{Q}^{\prime}\left(  R,0\right)   &  =-\int_{0}^{\infty}dT\,T\,\sigma
_{A_{b}}^{N}\left(  T|\left(  0,+\right)  _{A_{b}}\right)  =-R_{A_{b}}%
^{N}\left(  \left(  0,+\right)  _{A_{b}}\right)  T_{A_{b}}^{N}\left(  \left(
0,+\right)  _{A_{b}}\right) \\
\Psi^{\prime}\left(  0\right)   &  =-\int_{0}^{\infty}dT\,T\,\sigma_{A_{b}\cup
B}\left(  T|\left(  0,+\right)  _{A_{b}},\exists\tau<T:\lambda_{\tau}%
=\Delta\right) \nonumber\\
&  =-R_{A_{b}}^{B}\left(  \left(  \Delta,-\right)  _{A_{b}}\right)  \left\{
T_{A_{b}}^{B}\left(  \left(  \Delta,-\right)  _{A_{b}}\right)  +\frac{1}%
{\rho_{b}}+T_{A_{b}\cup B}\left(  \left(  \Delta,-\right)  _{A_{b}}\right)
\right\}
\end{align}
The timescales GG introduce are therefore%
\begin{align}
T_{d}  &  =-\frac{\tilde{\varPhi}^{\prime}\left(  d,0\right)  +\tilde
{Q}^{\prime}\left(  d,0\right)  }{\tilde{\varPhi}\left(  d,0\right)  }%
=\frac{R_{A_{c}}^{N}\left(  \left(  0,+\right)  _{A_{c}}\right)  T_{A_{c}}%
^{N}\left(  \left(  0,+\right)  _{A_{c}}\right)  +R_{A_{c}}^{C}\left(  \left(
0,+\right)  _{A_{c}}\right)  T_{A_{c}}^{C}\left(  \left(  0,+\right)  _{A_{c}%
}\right)  }{R_{A_{c}}^{C}\left(  \left(  0,+\right)  _{A_{c}}\right)  }\\
T_{R}  &  =-\frac{\tilde{Q}^{\prime}\left(  R,0\right)  +\Psi^{\prime}\left(
0\right)  }{\tilde{\varPhi}\left(  d,0\right)  }\nonumber\\
&  =\frac{R_{A_{b}}^{N}\left(  \left(  0,+\right)  _{A_{b}}\right)  T_{A_{b}%
}^{N}\left(  \left(  0,+\right)  _{A_{b}}\right)  +R_{A_{b}}^{B}\left(
\left(  \Delta,-\right)  _{A_{b}}\right)  \left\{  T_{A_{b}}^{B}\left(
\left(  \Delta,-\right)  _{A_{b}}\right)  +\frac{1}{\rho_{b}}+T_{A_{b}\cup
B}\left(  \left(  \Delta,-\right)  _{A_{b}}\right)  \right\}  }{R_{A_{c}}%
^{C}\left(  \left(  0,+\right)  _{A_{c}}\right)  }\nonumber\\
&  =\frac{T_{A_{b}\cup B}\left(  \left(  0,+\right)  _{A_{b}}\right)
}{R_{A_{c}}^{C}\left(  \left(  0,+\right)  _{A_{c}}\right)  }\\
T_{\nu}  &  =\frac{1}{\rho_{n}R_{A_{c}}^{C}\left(  \left(  0,+\right)
_{A_{c}}\right)  }%
\end{align}
so that, indeed, their expression%
\begin{equation}
\left\langle T\right\rangle =T_{d}+\frac{\left(  1-p_{c}\right)  }{p_{c}}%
T_{R}+\frac{1}{p_{c}}T_{\nu}%
\end{equation}
fully coincides with Eq.~(\ref{eq:Tsearch}).
\bibliographystyle{unsrt}
\bibliography{MTMFPT}

\begin{thebibliography}{10}

\bibitem{Alberts2002}
B.~Alberts, A.~Johnson, J.~Lewis, M.~Raff, K.~Roberts, and P.~Walter.
\newblock {\em {Molecular Biology of the Cell}}.
\newblock Garland Science, New York, 4th edition, 2002.

\bibitem{Mitchison1984}
Tim Mitchison and Marc Kirschner.
\newblock {Dynamic instability of microtubule growth}.
\newblock {\em Nature}, 312(5991):237--242, November 1984.

\bibitem{Dogterom1993}
Marileen Dogterom and Stanislas Leibler.
\newblock {Physical aspects of the growth and regulation of microtubule
  structures}.
\newblock {\em Physical Review Letters}, 70(9):1347--1350, March 1993.

\bibitem{Tran2001}
P~T Tran, L~Marsh, V~Doye, S~Inou\'{e}, and F~Chang.
\newblock {A mechanism for nuclear positioning in fission yeast based on
  microtubule pushing.}
\newblock {\em The Journal of cell biology}, 153(2):397--411, April 2001.

\bibitem{Grill2003}
Stephan~W Grill, Jonathon Howard, Erik Sch\"{a}ffer, Ernst H~K Stelzer, and
  Anthony~A Hyman.
\newblock {The distribution of active force generators controls mitotic spindle
  position.}
\newblock {\em Science (New York, N.Y.)}, 301(5632):518--21, July 2003.

\bibitem{Ambrose2011}
Chris Ambrose, Jun~F Allard, Eric~N Cytrynbaum, and Geoffrey~O Wasteneys.
\newblock {A CLASP-modulated cell edge barrier mechanism drives cell-wide
  cortical microtubule organization in Arabidopsis.}
\newblock {\em Nature communications}, 2:430, January 2011.

\bibitem{Holy1994}
T~E Holy and S~Leibler.
\newblock {Dynamic instability of microtubules as an efficient way to search in
  space.}
\newblock {\em Proceedings of the National Academy of Sciences of the United
  States of America}, 91(12):5682--5, June 1994.

\bibitem{Holy1997}
Timothy~Eric Holy.
\newblock {\em {Physical aspects of the assembly and function of
  microtubules}}.
\newblock PhD thesis, Princeton University, 1997.

\bibitem{Wollman2005}
R~Wollman, E~N Cytrynbaum, J~T Jones, T~Meyer, J~M Scholey, and A~Mogilner.
\newblock {Efficient chromosome capture requires a bias in the
  'search-and-capture' process during mitotic-spindle assembly.}
\newblock {\em Current biology : CB}, 15(9):828--32, May 2005.

\bibitem{Gopalakrishnan2011}
Manoj Gopalakrishnan and Bindu~S Govindan.
\newblock {A first-passage-time theory for search and capture of chromosomes by
  microtubules in mitosis.}
\newblock {\em Bulletin of mathematical biology}, 73(10):2483--506, October
  2011.

\bibitem{Redner2001}
S.~Redner.
\newblock {\em {A guide to first-passage processes}}.
\newblock Cambridge University Press, 2001.

\bibitem{Janson2003}
Marcel~E Janson, Mathilde~E de~Dood, and Marileen Dogterom.
\newblock {Dynamic instability of microtubules is regulated by force.}
\newblock {\em The Journal of cell biology}, 161(6):1029--34, June 2003.

\bibitem{Dogterom2005}
Marileen Dogterom, Jacob W~J Kerssemakers, Guillaume Romet-Lemonne, and
  Marcel~E Janson.
\newblock {Force generation by dynamic microtubules.}
\newblock {\em Current opinion in cell biology}, 17(1):67--74, February 2005.

\bibitem{Govindan2004}
Bindu Govindan and William Spillman.
\newblock {Steady states of a microtubule assembly in a confined geometry}.
\newblock {\em Physical Review E}, 70(3), September 2004.

\bibitem{Rubin1988}
R.~J. Rubin.
\newblock {Mean Lifetime of Microtubules Attached to Nucleating Sites}.
\newblock {\em Proceedings of the National Academy of Sciences},
  85(2):446--448, January 1988.

\bibitem{Williams1979}
David Williams.
\newblock {\em {Diffusions, Markov processes, and Martingales Volume I:
  Foundations}}.
\newblock John Wiley \& Sons, Chichester, August 1979.

\bibitem{Bicout1997}
D.~Bicout.
\newblock {Green's functions and first passage time distributions for dynamic
  instability of microtubules}.
\newblock {\em Physical Review E}, 56(6):6656--6667, December 1997.

\bibitem{Janson2007}
Marcel~E. Janson, Rose Loughlin, Isabelle Loïodice, Chuanhai Fu, Damian
  Brunner, François~J. Nédélec, and Phong~T. Tran.
\newblock Crosslinkers and motors organize dynamic microtubules to form stable
  bipolar arrays in fission yeast.
\newblock {\em Cell}, 128(2):357 -- 368, 2007.

\bibitem{Vos2004}
Jan~W. Vos, Marileen Dogterom, and Anne Mie~C. Emons.
\newblock Microtubules become more dynamic but not shorter during preprophase
  band formation: A possible "search-and-capture" mechanism for microtubule
  translocation.
\newblock {\em Cell Motility and the Cytoskeleton}, 57(4):246--258, 2004.

\end{thebibliography}

\end{document}